\documentclass[AMA,STIX1COL]{WileyNJD-v2}

\articletype{Article Type}%

\usepackage{amsmath}
\usepackage{graphicx,psfrag,epsf}
\usepackage{booktabs}
\usepackage{tabularx}
\usepackage{multirow}

\begin{document}

\newcolumntype{b}{X}
\newcolumntype{s}{>{\hsize=.5\hsize}X}

\title{Permutation-based multiple testing corrections for p-values and confidence intervals for cluster randomised trials}

\author[1]{Samuel I Watson*}

\author[2]{Joshua O. Akinyemi}

\author[1]{Karla Hemming}

\authormark{Samuel I Watson \textsc{et al}}

\address[1]{\orgdiv{Institute of Applied Health Research}, \orgname{University of Birmingham}, \orgaddress{\state{Birmingham}, \country{United Kingdom}}}

\address[2]{\orgdiv{College of Medicine}, \orgname{University of Ibadan}, \orgaddress{\state{Oyo State}, \country{Nigeria}}}


\corres{*Samuel I Watson, \email{s.i.watson@bham.ac.uk}}


\abstract[Summary]{In this article, we derive and compare methods to derive \textit{p}-values and sets of confidence intervals with strong control of the family-wise error rates and coverage for estimates of treatment effects in cluster randomised trials with multiple outcomes. There are few methods for \textit{p}-value corrections and deriving confidence intervals, limiting their application in this setting. We discuss the methods of Bonferroni, Holm, and Romano \& Wolf (2005) and adapt them to cluster randomised trial inference using permutation-based methods with different test statistics. We develop a novel search procedure for confidence set limits using permutation tests to produce a set of confidence intervals under each method of correction. We conduct a simulation-based study to compare family-wise error rates, coverage of confidence sets, and the efficiency of each procedure in comparison to no correction using both model-based standard errors and permutation tests. We show that the Romano-Wolf type procedure has nominal error rates and coverage under non-independent correlation structures and is more efficient than the other methods in a simulation-based study. We also compare results from the analysis of a real-world trial.}


\maketitle

\section{Introduction}
\label{sec:intro}

For a randomised controlled trial, the requirement to state a single primary outcome has become accepted, even required, practice. For example, the influential CONSORT statement on clinical trials requires the pre-specification of a single primary outcome, which they describe as the “outcome considered to be of greatest importance to relevant stakeholders”, and recommends against multiple primary outcomes.\citep{Schulz2010} The reason for this is to ensure appropriate control of the ``false discovery rate'' when using null hypothesis significance testing.\citep{Wason2014} If there are multiple outcomes each with their own associated treatment effect being tested separately, then we are implicitly testing a family of null hypotheses against an alternative that at least one of them is false. Without correction, the type I error rate for this family of null hypotheses will be much greater than the nominal rate of any single test.\citep{Romano2005} Indeed, the CONSORT statement notes that that multiple primary outcomes are not recommended as it ``incurs the problem of multiplicity of analyses''.\citep{TheLancet2010}

Cluster randomised trials are a widely used method to evaluate interventions applied to groups of people, such as clinics, schools, or villages. Often these interventions target `higher level' processes and can be complex in nature.\citep{Murray1998,Hemming2015b,Eldridge2012} Recent examples from our own work include an incentive scheme to improve implementation of a broad package of education and activities designed to improve employee health in the workplace,\citep{ThriveatWorkWellbeingProgrammeCollaboration2019} or a community health worker programme targeting multiple health conditions.\citep{Dunbar2018} The effects of such complex interventions cannot be adequately summarised by a single outcome. Creating a composite outcome is undesirable since it requires applications of arbitrary weights across outcomes and discards information by collapsing a multivariate outcome to a univariate one. The requirement for a single primary outcome therefore clashes with the needs of many cluster randomised trials. The solution is to ensure appropriate methods are used where there are multiple outcomes of interest rather than restricting the outcomes from which we can make inferences. However, the question of appropriate analysis for randomised trials, and particularly cluster randomised trials, with multiple outcomes can be contentious and complex. 

The Food and Drug Administration (FDA), the main regulatory body for medicines in the United States, declares that ``If the purpose of the trial is to demonstrate effects on all of the designated primary variables, then there is no need for adjustment of the type I error''.\citep{USFoodandDrugAdministration2017} They also identify a ``gatekeeping'' approach where ``statistical significance'' on a primary outcome is required before a second one can be analysed and state this does not need correction for multiple testing. Other authors differentiate aiming to declare ``statistical significance'' on at least one of a group of null hypotheses to requiring statistical significance for all tests in order to reject any individual test, and propose different solutions for both.\citep{LafayedeMicheaux2014,Rubin2021}

Where a correction for multiple testing is deemed necessary, we can divide solutions into: (i) multivariate methods that model the joint distribution of the outcomes, which is particularly favoured by Bayesian practitioners;\citep{Gelman2012} and (ii) univariate solutions that aim to ensure inferential statistics for a set of estimands collectively have the appropriate Frequentist properties.\citep{Farcomeni2008} In this article, we focus on the latter approaches in a Frequentist setting. Despite the different approaches and guidance, Wason \textit{et al}\citep{Wason2014} estimated that only around half of all randomised trials with multiple outcomes or arms corrected for multiple testing. No evidence is available on the use of corrections for multiple testing in cluster randomised trials specifically, but there are few, if any, comprehensive discussions of methods in this area currently available. Furthermore, almost all discussion of multiple testing adjustment relates to corrections for p-values, with few, if any, solutions for confidence intervals. The FDA note that correcting confidence intervals is complex and beyond the scope of their advice. However, the duality between hypothesis testing and confidence intervals means that we should be able to identify the bounds of a `confidence set' adjusted for multiple testing.\citep{Romano2005, Romano2005a} The primary limiting factor to using corrected confidence intervals is that there are no proposed methods for determining these bounds efficiently. 

In this article, we develop several methods for adjusting p-values for multiple testing for a cluster randomised trial setting using permutation-test based methods, by adapting existing methods of correction, and propose a novel method to derive corrected confidence sets. We then compare these methods in a simulation-based study to evaluate type I error rates and efficiency of the different procedures. Our analysis is based on generalised linear mixed models, which are frequently used in the analysis of cluster trials. We also focus on permutation-based methods, since these methods provide exact inference at all sample sizes. A small number of clusters, which is common to many cluster trials, can result in small sample biases in the standard error estimator and inflated type 1 errors,\citep{Watson2021,Leyrat2018,Li2017} which results in complication when it comes to considering additional corrections for multiple testing. Section 2 provides a review and discussion of methods for correcting for multiple testing and their adaptation to a cluster randomised trials setting, Section 3 presents a simulation-based comparison, Section 4 provides an applied example, and Section 5 concludes.


\section{Multiple testing in cluster randomised trials}
\subsection{The multiple testing problem}
We first suppose that data $\mathbf{X}$ are generated from some probability distribution $P$, which belongs to some family of probability distributions $\Omega$. The family $\Omega$ could be a parametric, semi-parametric, or non-parametric model. The multiple testing problem arises when we have a set of hypotheses $H_j$ versus $H_{j'}$ for $j=1,...,J$, following the notation of Romano and Wolf.\citep{Romano2005} These hypotheses in our context are typically estimates of the treatment effect of an intervention on multiple outcomes. Each of the hypotheses is a subset $\omega_j \subset \Omega$ and is equivalent to testing $P\in \omega_j$ against $P \not\in \omega_j$. So for any subset $K \subset {1,...,J}$, $H_K = \cap_{j\in K}H_j$ is the hypothesis that $P \in \cap_{j \in K} \omega_j$. We assume each null hypothesis $H_j$ is based on a test statistic $T_j$; we denote the $\alpha$-quantile of the distribution of $T_j$ as $c_j(\alpha,P)$. In a traditional null hypothesis testing framework we ``reject'' $H_j$ in favour of $H_{j'}$ at the $\alpha$ level, if $T_j \geq c_j(1-\alpha,P)$, which clearly has probability $\alpha$. Conversely, the \textit{p}-value $p_j$ of the test is where $T_j = c_j(1-p_j,P)$, so that the probability of observing $Pr(T_j > c_j(1-p_j,P)|H_j) = p_j$. The family-wise error rate (FWER) of this set of hypotheses is the probability of ``rejecting'' at least one true null hypothesis. That is, if $I = I(P) \subset {1,...,J}$ are the indices of the true null hypotheses, so $j \in I$ if and only if $P \in \omega_j$, then the FWER is the probability under $P$ of rejecting any $H_{j\in I}$, i.e. $Pr(\cup_{j\in I}T_j > c_j(1-\alpha,P))$, which should be $\alpha$. 

\subsection{Methods for correcting for multiple testing}
\label{sec:perms}
Solutions to the multiple testing problem aim to ensure that $FWER \leq \alpha$. Control over the FWER is said to be \textit{strong} if it holds for any combination of true and false null hypotheses, and \textit{weak} if it only holds when all null hypotheses are true.\citep{Dudoit2003} Several approaches exist to control the FWER. The Bonferroni method is probably most well known, which sets the critical value for the test of the null hypothesis to be $c_j(1-\alpha/J,P)$. Equivalently, \textit{p}-values that maintain the FWER for the family of null hypotheses ensure that $Pr(\cup_{j\in I}T_j > c_j(1-p,P)) = p$, so a crude `corrected' \textit{p}-value for the null hypothesis $H_j$ using the Bonferroni method would be $\min(Jp_j,1)$. However, while this method exerts strong control over the FWER, it is highly conservative.

Holm\citep{Holm1979} proposed a less conservative `stepdown' approach to multiple testing. One orders the test statistics from largest to smallest and then compares the largest statistic to the critical value $c_j(1-\alpha/J,P)$. If the test statistic is larger than this value, then the null hypothesis is rejected, otherwise we do not reject any null hypothesis and stop. If we rejected, then the next largest test statistic is compared to $c_j(1-\alpha/(J-1),P)$, and again it is either rejected, or we do not reject all remaining null hypotheses and stop, and so forth. A crude corrected \textit{p}-value could therefore be obtained by multiplying the smallest to the largest \textit{p}-values by $J$, $J-1$, etc, respectively. The Holm method is less conservative than the Bonferroni method,\citep{Holm1979} but it may still be inefficient as, like the Bonferroni method, it does not explicitly take into account the dependence structure in the data. Romano and Wolf\citep{Romano2005, Romano2005a} developed an efficient resampling based version of Holm's stepdown method, which can use permutation-based tests in the context of a cluster randomised trial.

\subsection{Permutation-based corrections for multiple testing}
An issue that complicates analyses of cluster randomised trials is that test statistics can fail to have the expected sampling distribution in a range of circumstances, but particularly when the number of clusters is small.\citep{McNeish2016,Watson2021,Li2017,Leyrat2018} This issue means determining the critical value of a hypothesis test, even in the absence of any multiple testing issue, can be difficult. While there exist several small sample corrections in the literature their performance often depends on the correlation structure, which is not known.\citep{Watson2021,Li2017} 

An alternative approach is to use a permutation testing method based on the randomisation scheme for the trial. In particular, the null hypothesis implies that the distribution of the data $\mathbf{X}$ is invariant under a set of transformations in $\mathbf{A}$, which has $L$ elements. So, $a\mathbf{X}$ and $\mathbf{X}$ have the same distribution for all $a \in \mathbf{A}$ whenever $\mathbf{X}$ has distribution $P \in \omega$. $\mathbf{A}$ in the context of cluster randomised trials is the set of all transformations that could be generated by the randomisation mechanisms, for example, all ways of dividing the clusters into two groups for a parallel design. Our observed test statistics with our sample data are $T_j(\mathbf{X})$. The test statistic generated by the $l$th permutation is $T_j(a_l\mathbf{X})$ for $a_l \in \mathbf{A}$ and $l=1,...,L$. We can use this approach to estimate the critical values for the Bonferroni or Holm corrections. For example, for Bonferroni:
\begin{equation}
    \hat{c}_{j}(1-\alpha/J,P) = T_{j,|L(1-\alpha/J)|}
\end{equation}
where $T_{j,|L(1-\alpha/J)|}$ is the $L(1-\alpha/J)$th (or nearest integer) largest value from the permutations. And a crude, corrected two-sided \textit{p}-value is:
\begin{equation}
\label{eq:pvalue}
    p_j = \min \left( \frac{J}{L}  \sum_{l=1}^L \mathbf{1}\left[ \text{abs}(T_j(a_l\mathbf{X})) \geq \text{abs}(T_j(\mathbf{X})) \right] , 1\right)
\end{equation}
where $\mathbf{1}$ is the indicator function and $\text{abs}$ is the absolute value. The same approach can be used for the Holm method. 

Romano and Wolf\citep{Romano2005, Romano2005a} developed a modified stepdown approach to take advantage of resampling methods. Their process is optimal in a maximin sense. We describe the general stepdown procedure of Romano and Wolf firstly in terms of accepting or rejecting each null hypothesis at an $\alpha$-level. We let $c_{\mathbf{K}}(\alpha,P)$ denote an $\alpha$-quantile of the distribution of the statistic:
\begin{equation}
\label{eq:maxstat}
    T_K = \text{max}_{j\in K}T_j
\end{equation}
for any subset of null hypotheses $K$. We also denote $T_{|r|}$ as the $r$th largest test statistic so that
\begin{equation}
    T_{|1|} \geq T_{|2|} \geq ... \geq T_{|J|}
\end{equation}
corresponding to hypotheses $H_{|1|}$, $H_{|2|}$, ..., $H_{|J|}$. Then the idealised algorithm is:
\begin{enumerate}
    \item Let $K_1={1,...,J}$. If $T_{|1|} \leq c_{K_1}(1-\alpha,P)$ then accept all hypotheses and stop; otherwise, reject $H_{|1|}$ and continue;
    \item Let $K_2$ be the indices of all the hypotheses not previously rejected. If $T_{|2|} \leq c_{K_2}(1-\alpha,P)$, then accept all remaining hypotheses and stop; otherwise, reject $H_{|2|}$ and continue;
    \item[{$\vdots$}] 
    \item[$J$.] If $T_{|J|} \leq c_{K_J}(1-\alpha,P)$ then do not reject $H_{|J|}$, otherwise reject.
\end{enumerate}
In this procedure it is assumed the critical values are known. One can see that this algorithm replicates Holm's procedure, but allows us to use permuation-based methods to estimate the critical values where they are not known.

 For each permutation we can determine the test statistic as in Equation (\ref{eq:maxstat}) as $T_{K,l} = \text{max}_{j\in K}T_j(a_l\mathbf{X})$. As before we denote $T_{K,|r|}$ as the $r$th largest of all the permutational test statistics $\{T_{K,l};l=1,...,L\}$. Then our estimator for the critical value is:
\begin{equation}
    \hat{c}_{K}(1-\alpha,P) = T_{K,|L(1-\alpha)|}
\end{equation}
 We can see how this procedure produces \textit{p}-values for a two-sided hypothesis that also maintains the FWER for a given $\alpha$ \citep{Romano2016}, in particular:
\begin{equation}
\label{eq:pvalue2}
    p_K = \frac{1}{L}  \sum_{l=1}^L \mathbf{1}\left[ \text{abs}(T_K(a_l\mathbf{X})) \geq \text{abs}(T_K(\mathbf{X})) \right] 
\end{equation}
For a one-sided test we would not use the absolute values of the test statistics.

Often the size of $\mathbf{A}$ can be very large, and increases exponentially with the number of clusters. A Monte Carlo approach can be used that instead generates a random subset of $\mathbf{A}$ of fixed sized in order to generate realisations of the test statistics. If we conduct $M$ such permutations then the estimator of the \textit{p}-value for a given null hypothesis versus some alternative is
\begin{equation}
    \hat{p}_K = \frac{1}{M+1}\sum_{m=1}^M \left( 1+ \mathbf{1}[\text{abs}(T(a_m\mathbf{X})) \leq \text{abs}(T(\mathbf{x})) ] \right)
\end{equation}
Obtaining \textit{p}-values in this way is described in detail by Romano.\citep{Romano2016} Values of $M=1,000$ or greater are often used as this results in relatively small Monte Carlo error, although much larger values (e.g. 10,000 or 100,000) may be preferred for formal or final analyses.

In subsequent sections, we develop and compare Bonferroni, Holm, and Romano-Wolf methods, however, we note there are several other multiple testing corrections in the literature, including Hochberg's `step-up' procedure,\citep{HOCHBERG1988} Hommel's `stagewise' procedure,\citep{HOMMEL1988} and Šidák's procedures\citep{Holland1987} (see also \citep{Farcomeni2008} for a discussion). More exhaustive comparisons of these methods in other settings, such as\citep{Stevens2017,Vickerstaff2019,Ozenne2022,Alberton2020}, show that they all maintain a FWER $\leq \alpha$, but that Holm's, Hommel's, and Hochberg's procedures generally are the most efficient and perform very similarly. However, these comparisons do not include the Romano-Wolf method, which purports to be at least as efficient as Holm's procedure.\citep{Romano2005} We note that Westfall and Young\citep{Westfall1993} propose an early version of a resampling based multiple testing correction similar to Romano-Wolf, which is included in the comparison by Alberton et al\citep{Alberton2020} in the context of modelling brain imaging data. We adapt only a subset of all methods, but believe the application of other methods in the context we describe below, including any developed after the publication of this article, should be clear from the discussion of these four key approaches. 
 
\subsection{Permutation test statistics for cluster trials}
We next introduce a generalised linear mixed model commonly used in the analysis of cluster randomised trials (e.g.\citep{Hooper2016}). We denote $Y_{ict}$ as the outcome of the $i$th individual, $i=1,...,N$, in cluster $c=1,...,C$ at time $t=1,...,T$. We include a temporal dimension in this discussion for generality, however, it can be ignored as required. Our simulation-base comparisons include both examples with and without a temporal dimension. We do not restrict the outcome, it could be continuous or discrete. We specify the linear predictor:
\begin{equation}
\label{eq:linpred}
    \eta_{ict} = \mu_0 + \delta D_{ct} + X_{ict}'\beta + \theta_{ct}
\end{equation}
where $D_{ct}$ is an indicator for whether cluster $c$ has received the intervention at time $t$ and so $\delta$ is the parameter of interest, our ``treatment effect''. We also have a vector of individual and/or cluster-level covariates, $X_{ict}$, which may also contain temporal fixed effects. The parameter $\theta_{ct}$ represents a general `random-effect' term that captures the within cluster and cluster-time correlation, although we do not provide a specific structure here. The overall model is then 
\begin{equation}
\label{eq:crtmod}
    Y_{ict} \sim P(h(\eta_{ict}))
\end{equation}
where $h(.)$ is a link function. For example, $P$ could be a Binomial distribution and $h(.)$ the logistic link function. 

Gail \textit{et al}\citep{Gail1996a} provided the first extensive examination of permutation tests for cluster-based study designs. Their work principally used unweighted differences of cluster means as the basis of permutation tests (see also \citep{Thompson2019}). Several other authors have also developed and evaluated permuatation-tests and test statistics in the context of cluster trials.\citep{Wang2017,Zhou2022,Murray2006,Li2016,Li2022,Blaha2022} Here, we build on the statistic proposed by Braun and Feng\citep{Braun2001}.

Braun and Feng\citep{Braun2001} examine optimal permutation tests for cluster randomised trials specifically. They derive a `quasi-score' statistic using the marginal likelihood of the data modelled separately from the correlation structure of the data. The marginal mean of each observation, ignoring the cluster-effects $\theta_{ct}$, is 
\begin{equation}
\label{eq:meanglm}
    h^{-1}(\mu_{ict}) = \mu_0 + \delta D_{ct} + X_{ict}'\beta
\end{equation}
The ``quasi-score'' statistic, which is weighted sum of generalised residuals, is then:
\begin{equation}
\label{eq:teststat}
    \sum_{c} \{ D^*_{c}\mathbf{G}_{c}\mathbf{V}_{c}^{-1}[\mathbf{Y}_c - \boldsymbol{\mu}_{c}] \} |_{\delta = \delta^*}
\end{equation}
where $D^*_c [D^*_{c1},D^*_{c1},D^*_{c1},...,D^*_{cT},D^*_{cT}]'$ is a $(1 \times n_{c})$ vector of modified intervention indicators equal to 1 if the intervention was present in cluster $c$ at time $t$ and -1 otherwise, and where $n_c = \sum_{t}n_{ct}$ and $n_{ct}$ is the number of individuals in cluster $c$ at time $t$. $\mathbf{G}_c$ is a $(1 \times n_{c})$ vector with elements $(\partial h_{ict}^{-1}/\partial \eta_{ict})^{-1}$, and $\mathbf{V}_c$ is an $(n_c \times n_{c})$ covariance matrix for cluster $c$ with non-zero elements off its diagonal. As an example, if we assume the data are normally distributed with mean $\mu_{ict}$, identity link function, variance $\sigma^2$, and $\theta_{ct} \sim N(0,\tau^2)$, then the diagonal elements of $\mathbf{V}_c$ are $\sigma^2 + \tau^2$ and the off-diagonal elements are $\tau^2$. More complex structures might include temporal decay in correlation, for example. We use $\Theta$ to represent the parameters of the variance-coviarance matrix. Finally $[\mathbf{Y}_c - \boldsymbol{\mu}_{c}]$ are generalised residuals: $\mathbf{Y}_c = [Y_{1c1},Y_{2c},...,Y_{n_{c1}c1},Y_{1c2},...,Y_{n_{cT}cT}]$ is a $(1 \times n_{c})$ vector of outcomes and $\boldsymbol{\mu}_{c}$ is a $(1 \times n_{c})$ vector of means. 

For the permutation test to be valid the `nuisance' parameters $(\mu, \beta, \Theta)$, i.e. those other than $\delta$, must be invariant to permutation.\citep{Braun2001} This means we cannot re-estimate them for each new permutation. In practice the maximum likelihood estimates of these parameters are used to construct the test statistic, so that we use the estimates:
\begin{equation}
\label{eq:meanglm2}
    \hat{\mu}_{ict} = h(\hat{\mu}_0 + \delta^* D_{ct} + X_{ict}'\hat{\beta})
\end{equation}
for the linear predictor under the null $H_0:\delta=\delta^*$. Estimating $\Theta$ is more difficult, however, particularly when the number of clusters is small.\citep{McNeish2016,Watson2021} As an alternative to (\ref{eq:teststat}) we can replace $\mathbf{G}_{c}\mathbf{V}_{c}^{-1}$ with a $(1 \times n_c)$ vector of ones:
\begin{equation}
\label{eq:teststat2}
    \sum_{c} \sum_t \sum_i \{ D^*_{ict} [Y_{ict} - \mu_{ict}] \} |_{\delta = \delta^*}
\end{equation}
so that the sum of residuals is `weighted' only by the size of each cluster or cluster-time period. One can see that under homoscedasticity the two test statistics will be approximately proportional. The weighted statistic weights the residuals in proportion to their variance, so in non-linear models with differing variances (e.g. different linear predictors over time) we may expect to see an improvement in efficiency.

The quasi-score statistics are the motivation behind quasi-likelihood approaches, including GEE methods.\citep{LIANG1986, Braun2001} Thus, the tests and corrections described here can be implemented within a GEE framework. However, our simulations in Section 3, model estimation, and the software we provide to implement the methods uses a more explicitly GLMM formulation. The quasi-score statistic is equivalent for full and marginal likelihoods using linear Gaussian models, or when using the `unweighted' variant described below. For non-linear alternatives though, the quasi-score statistic is an approximation to the full likelihood. In terms of our implementation of the computation of (\ref{eq:teststat}), we use a GLMM formulation (Equation \ref{eq:linpred}) and use the original estimates of the covariance parameters to generate an estimated inverse covariance matrix $\hat{\mathbf{V}}^{-1}$, which is then re-used for each iteration. 

For the purposes of correcting for multiple testing we use studentized versions of the two test statistics:
\begin{align}
T_w = T_w(\mathbf{X})|_{\delta = \delta_0} &= \frac{\sum_{c} \{ D^*_{c}\mathbf{G}_{c}\mathbf{V}_{c}^{-1}[\mathbf{Y}_c - \boldsymbol{\mu}_{c}] \}}{\sqrt{\sum_{c} \{ D^*_{c}\mathbf{G}_{c}\mathbf{V}_{c}^{-1}[\mathbf{Y}_c - \boldsymbol{\mu}_{c}] \}^2}} \\
    T_u = T_u(\mathbf{X})|_{\delta = \delta_0} &= \frac{\sum_{c} \sum_t \sum_i \{ D^*_{ict}\mathbf{1}_{ict}[Y_{ict} - \mu_{ict}] \}}{\sqrt{\sum_{c} \sum_t \sum_i \{ D^*_{ict}\mathbf{1}_{ict}[Y_{ict} - \mu_{ict}] \}^2}}
\end{align}
where the terms on the right-hand side have been evaluated at $\delta = \delta^*$. We describe $T_w$ as the ``weighted test statistic'' and $T_u$ as ``unweighted''. In the absence of studentization, the variances of the test statistics are not scale-free and depend on, among other things, the null hypothesis being tested so that different tests will have different power.\citep{Romano2005} The lack of balance is particularly consequential for the construction of confidence sets discussed in the next section. While confidence sets constructed on the basis of permutational methods will have joint coverage of $1-\alpha$, without balance the individual coverage probabilities of each interval will differ, perhaps substantially.\citep{Romano2005a}


\subsection{Confidence sets and multiple testing}
The multiple testing problem extends to the construction of simultaneous confidence intervals or a ``confidence set''. Let the parameters of interest be $\delta_j$ with associated confidence intervals $[L_j, U_j]$, so that $[L_1, U_1] \times [L_2, U_2]$   $... \times [L_J, U_J]$, $\mathbf{U} = [U_1,...,U_J]$ and $\mathbf{L} = [L_1,...,L_J]$, forms a confidence set. Similar to the FWER, we want appropriate control of the coverage of the $100(1-\alpha)\%$ confidence set such that the process produces confidence sets with the property:
\begin{equation}
\label{eq:ciprob}
    Pr(\cup_j \delta_j \in [L_j, U_j]) = 1-\alpha
\end{equation}
we refer to this as `family-wise coverage', which we use analogously to `simulatanous coverage' used in other contexts. If we construct $100(1-\alpha)\%$ confidence intervals independently then the probability that at least one interval in the set excludes the true value can significantly exceed $\alpha$. For Bonferroni, an obvious modification is to instead estimate $100(1-\alpha/J)\%$ confidence intervals to acheive a family-wise coverage of $100(1-\alpha)\%$. There have been some attempts to construct exact confidence sets for parameters analytically based on the stepdown procedure.\citep{Romano2005a} For example, Guilbaud,\citep{Guilbaud2008} extending the proposal of Hayter and Hsu,\citep{Hayter1994} uses the acceptance/rejection of null hypotheses by the stepdown procedure as a basis of determining upper or lower limits of confidence intervals if we conclude they are strictly negative or postitive, respectively. However, these procedures can only provide information on the upper or lower bound respectively - the other end of the interval is infinity - so they provide little extra information on the extent of sampling variation beyond the \textit{p}-value. 

As an alternative, consider for a moment, a single parameter $\delta_1$. Its $100(1-\alpha)$\% confidence interval is $[L_1,U_1]$: for any value $\delta^*_1$ inside this interval the null hypothesis $H_1:\delta_1 = \delta^*_1$ will not be rejected in favour of the two-sided alternative $H_{1'}: \delta_1 \neq \delta^*_1$ at the $\alpha$ level. The question is then how to find the values of $L_1$ and $U_1$ efficiently. One could iteratively perform a series of permutation tests to identify the limits as $U_1 = \text{sup}\{ \delta_1^*: \text{do not reject } \delta_1 = \delta_1^* \}$ and $L_1 = \text{inf}\{\delta_1^*: \text{do not reject } \delta_1 = \delta_1^* \}$. However, this procedure is inefficient, particularly when testing multiple parameters: if there are $M$ permutations per test and $J$ outcomes, then for each increment in $\mathbf{U}$ we must calculate $JM$ permutation test statistics and perform the desired correction. Moreover, since the test statistic and its permutational distribution depends on the values of the other null hypotheses being tested, a very large number of combinations of values of the parameters must be tested to ensure we have identified with reasonable certainty the limits of the confidence set.

Garthwaite and Buckland\citep{Garthwaite1992} developed a method for searching for confidence interval endpoints efficiently, which Garthwaite\citep{Garthwaite1996} later adapted for use with permutation tests. Their method is based on the search process devised by Robbins and Munro,\citep{Robbins1951} who developed a stochastic approximation procedure to find the $\alpha$-quantile of a particular distribution. Multivariate Robbins-Monro processes follow the same procedures as their univariate equivalents.\citep{Ruppert1985} For our multiple testing scenario the upper limits to the confidence set correspond to where all hypotheses $H_j: \theta_j = U_j$ for $j=1,...,J$ are all rejected in favour of the two-sided alternative with a FWER of $\alpha$ but for any smaller values of $U_j$ not all hypotheses are rejected, and equivalently for the lower limits. Rabideau et al\citep{Rabideau2021,Rabideau2021b} have also independently proposed this method for confidence interval estimation for cluster randomised trials, although not in the context of multiple testing.

For each method, at the $q$th step of $Q$ steps total, we have estimates of the upper confidence interval limits of our $J$ parameters $\mathbf{u}_q = [u_{1q}, u_{2q},...,u_{Jq}]$. We generate the set of test statistics $T_j(\mathbf{X})|_{\delta=u_{jq}}$, which correspond to the null hypotheses $H_j: \delta_j = u_{jq}$. We then generate a single permutation of a permutation test for the same hypotheses $\text{abs}(T_j(a_q\mathbf{X}))_{\delta=u_{jq}}$. Each method then defines a procedure for determining whether to reject these hypotheses or not, which are described in the preceding sections. For example, with the Romano-Wolf stepdown procedure: reject hypothesis $H_{|1|}$ if $\text{abs}(T_{K_1}(a_q\mathbf{X})) < \text{abs}(T_{|1|}(\mathbf{X}))$ otherwise do not reject any hypothesis and stop; if $H_{|1|}$ was rejected then reject hypothesis $H_{|2|}$ if $\text{abs}(T_{K_2}(a_q\mathbf{X})) < \text{abs}(T_{|2|}(\mathbf{X}))$ otherwise do not reject any further hypotheses and stop, and so forth. 

The estimates of the upper limits are updated based on the single permutation draw as (we drop the subscript $\delta=u_{jq}$ for ease of notation, but the test statistics are evaluated at this value for each iteration):
\begin{equation}
    u_{j,q+1} = 
    \begin{cases}
    u_{jq} - s_j \alpha^*/q &\text{if   } H_j \text{ rejected} \\
    u_{jq} + s_j(1-\alpha^*)/q &\text{otherwise}
    \end{cases}
\end{equation}
where $s_j$ is the ``step length constant''. With no correction and with Romano-Wolf $\alpha^* = \alpha$, for Bonferroni $\alpha^* = \alpha/J$, and for Holm $\alpha^* = \alpha/J$ for $H_{|1|}$, $\alpha^* = \alpha/(J-1)$ for $H_{|2|}$, and so forth. Similarly for the lower limits, the updating rule is:
\begin{equation}
    l_{j,q+1} = 
    \begin{cases}
    l_{jq} + s_j\alpha^*/q &\text{if } H_j \text{ rejected}\\
    l_{jq} - s_j(1-\alpha^*)/q &\text{otherwise}
    \end{cases}
\end{equation}
The step length constants are $s_j = k(u_{jq}- \hat{\theta}_j)$ and $s_j = k(\hat{\theta}_j - l_{jq})$ for the upper and lower limits, respectively, where $\hat{\theta}_j$ is a point estimate of the parameter and:
\begin{equation}
    k = \frac{2}{z_{1-\alpha}(2\pi)^{-1/2}\text{exp}(-z_{1-\alpha}^2/2)}
\end{equation}
where $z_\alpha$ is the $\alpha$-quantile of the standard normal distribution. The algorithm proceeds for a pre-selected number of iterations; in the simulations in the subsequent section we have used 2,000 iterations. A sensible starting value for this algorithm is the approximate uncorrected confidence interval limits, for example, for the upper limit $u_j,0 = \hat{\beta}_j + 2SE_j$ where $SE_j$ is the standard error of $\beta_j$ from the univariate model.


\subsection{Computation}
An R package developed by the authors to execute the analyses described in this paper is available from CRAN as \texttt{crctStepdown} (version 0.2.1 at the time of writing) including implementations of the Romano-Wolf, Holm, and Bonferroni methods for correcting p-values and confidence sets using permutation-based tests.

\section{Simulation study}
\subsection{Methods}
We conduct a simulation-based study to examine the FWER, family-wise coverage, and efficiency of the procedures outlined in the previous sections for cluster randomised trials. We compare the following procedures:
\begin{enumerate}
    \item A `naive' no correction approach using the reported standard errors and test statistics from the output of the \texttt{lme4} package for R. 95\% confidence intervals for each parameter were constructed as $\hat{\delta} +/- 1.96 \times SE$.
    \item No correction with p-values and confidence sets derived from permutation based tests.
    \item The Bonferroni method using permutation based tests.
    \item The Holm method using permutation based tests.
    \item The Romano-Wolf method using permutation based tests.
\end{enumerate}
For methods 2-5 we use both the weighted and unweighted test statistic resulting in nine methods. For the Bonferroni and Holm methods we only use permutation-based inference rather than the perhaps more standard approach of adjusting p-values reported by mixed model fitting software. Model-based inference can fail to have nominal FWERs for reasons other than multiple testing, such as biases arising from small numbers of clusters, which would further complicate interpretation of the results. We include a comparison with methods 1 and 2 to illustrate this issue in our context.

\subsubsection{Data generating processes}
We use three different data generating processes of cluster randomised trials, described below. We opt for specific scenarios of rising complexity to examine the performance of the nine different methods (including both unweighted and weighed versions of the permutation-based methods). All outcomes are simulated and modelled using exponential-family models. In all simulations we set the number of individuals per cluster to 20 and simulate either seven or 14 clusters per arm. The choice of number of clusters is informed by two key considerations. First, the simulations take a very long time to run given the number of GLM models required to be estimated for the permutation tests and search procedures (for three outcomes and 10,000 iterations we require 90 million models), and so we aimed to choose the smallest number that would provide the desired inference. Second, we wanted to include scenarios where there was likely small sample bias in `standard' non-permutation based estimators of standard errors due to the low number of clusters, and one where such biases were likely minimal. Previous literature on cluster trials suggests small sample biases are likely minimal at 14 clusters or more per arm, but present with seven clusters per arm [REF], although permutation-based methods provide exact inference at any sample size. We provide estimates of FWER without correction and with non-permutation based estimators to examine whether there are likely small sample biases. However, we recognise that 14 clusters per arm may still be considered `small'. The treatment effect parameters for each simulation are a vector, $\delta$, with length equal to the number of outcomes and with different combinations of either 0 or 1, allowing for when all treatment effects are zero and when only a subset are.

\textbf{(1) Two-arm, parallel cRCT, two outcomes}

The first simulation data generating process (`model (1)') represents a two arm parallel cluster trial with two outcomes measured once in the post-intervention period. Both outcomes $Y_j$ are continuous, Gaussian variables for $j = 1,2$. This model is intended to examine the effect of correlation, which we model at the individual and cluster levels. For individual $i$ in cluster $c$:
\begin{equation}
    \begin{pmatrix}
    Y_{1,ic} \\
    Y_{2,ic}
    \end{pmatrix} \sim 
    \left(  \begin{pmatrix}
    \mu_1 + \delta_1 D_{c} + \theta_{1,c} \\
    \mu_2 + \delta_2 D_{c} + \theta_{2,c}
    \end{pmatrix},
     \begin{pmatrix}
    \sigma^2_1 & \rho \sigma_1 \sigma_2 \\
    \rho \sigma_1 \sigma_2 & \sigma^2_2
    \end{pmatrix} \right)
\end{equation}
where $\mu_j$ are intercept parameters, $D_c$ is an indicator for whether the cluster is treated or not, and $\theta_{j,c}$ are cluster level random effect modelled as:
\begin{equation}
\label{eq:randeff}
    \begin{pmatrix}
    \theta_{1,c} \\
    \theta_{2,c}
    \end{pmatrix} \sim 
    \left(  \begin{pmatrix}
    0 \\
    0
    \end{pmatrix},
     \begin{pmatrix}
    \tau^2_1 & \pi \tau_1 \tau_2 \\
    \pi \tau_1 \tau_2 & \tau^2_2
    \end{pmatrix} \right)
\end{equation}
The parameters $\rho$ and $\pi$ are correlation parameters at the individual and cluster levels, respectively with $\sigma_j$ and $\tau_j$ the standard deviation of the individual-level outcomes and cluster-random effect terms, respectively. Clusters are assigned in a 1:1 ratio with seven or 14 clusters per arm and 20 or 10 individuals per cluster. We set $\mu_j = 1$ and consider both $\delta = (0,0)$ and $\delta = (0,0.5)$ to compare the FWER under different combinations of true null hypotheses. We set $\sigma^2_j = 1$ and $\tau^2_j = 0.05$, which gives a marginal intraclass correlation coefficient (ICC) ($\text{ICC}_j = Var(\theta_{j,c})/Var(Y_{j,ic})$) of 0.05. We also set $\rho = \pi$ and examine a range of values. We do not report outcomes using the weighted test statistic with this example as it is proportional the unweighted test statistic as both models are Gaussian with identity link, so there will be no difference in performance. 

\textbf{(2) Two-arm, parallel cRCT, two differently distributed outcomes}

For the next set of simulations (`model (2)') we consider a parallel cluster trial with two outcomes measured once in the post-intervention period. Simulation parameters are as the previous example, unless stated below. The first outcome is specified as Poisson distributed:
\begin{equation*}
    Y_{1,ic} \sim \text{Poisson}(\text{exp}(\mu_1 + \delta_1 D_{c} + \theta_{1,c})) 
\end{equation*}
and the second outcome as Gaussian distributed:
\begin{equation*}
    Y_{2,ic} \sim N(\mu_2 + \delta_2 D_{c} + \theta_{2,c}, 1)
\end{equation*}
where the random effects are specified as in Equation (\ref{eq:randeff}) with $\pi = 0$. We again set $\mu_j = 1$ and consider both $\delta = (0,0)$ and $\delta = (0,0.5)$. The ICC for non-linear models depends on the realised values of the covariates and the parameter values and so will differ between simulations. We again choose $\sigma_j^2 = 0.05$, which gives a range of ICCs between approximately 0.01 and 0.2 for the Poisson model and 0.05 for the Gaussian model. 



\textbf{(3) Two-arm parallel cRCT with baseline measures, three outcomes}

We finally extend the parallel cluster trial model (`model (3)') to include baseline measures, which incorporates a temporal dimension and hence more complex covariance structure. The trial includes seven clusters in each arm, with half receiving the intervention in the second time period. We simulate three outcomes, with index $t$ representing time period:
\begin{align}
\begin{split}
    Y_{1,ict} &\sim \text{Poisson}(\text{exp}(\mu_1 + \delta_1 D_{ct} + T_1 + \theta_{1,ct})) \\
    Y_{2,ict} &\sim N(\mu_2 + \delta_2 D_{ct} + T_2 + \theta_{2,ct}, 1)\\
    Y_{3,ict} &\sim \text{Bernoulli}(\text{logit}(\mu_3 + \delta_3 D_{ct} + T_3 + \theta_{3,ct})) 
    \end{split}
\end{align}
where, now, $D_{ct}$ equals one if the cluster has the intervention in time period $t$ and zero otherwise and $T$ is a fixed effect for the second time period. We use an auto-regressive specification for $\theta_{j,ct}$ to facilitate incorporation of correlation between outcomes. In particular,
\begin{align}
    Cov(\theta_{j,ct},\theta_{j,ct'}) &= \lambda^{\vert t - t' \vert} \sigma^2_j \\
    Cov(\theta_{j,ct},\theta_{j',ct'}) &= \lambda^{\vert t - t' \vert} \sigma_j \sigma_{j'} \rho 
\end{align}
for $j \neq j'$. The random effects have a multivariate normal specification as before zero correlation. We maintain the same number of individuals per cluster. We set $\mu_j = -1$ and $\tau_j = 1$ for all $j=1,2,3$. We vary the choice of $\delta$ as either $(0,0,0)$ or $(0,0.5,0)$; as with the previous set of simulations we do not consider a completely exhaustive set of permutations of simulation parameters. We set $\lambda = 0.7$.

\subsubsection{Simulation methods}
Each set of simulations is run 10,000 times. We note the Monte Carlo error will be moderately higher than expected due to variation arising from the permutation tests, confidence set search procedure, and simulations. We use 1,000 iterations for the permutation test \textit{p}-values and 2,000 steps for the search procedure as these produced stable values for these simulations (although we note that for more outcomes longer runs were often required for the confidence interval search procedure for it to reach a stable equilibrium). Point estimates of parameters were obtained from univariate generalised linear mixed models estimated with the R package \texttt{lme4} for models (1) and (2), we similarly obtained estimates of variance parameters from these models for the weighted test statistics. For example (3) we obtained parameter estimates from a generalised linear model with no random effects given the lack of widely available software for estimating autoregressive random effects models; weighted test statistics were generated using a covariance matrix created with the values of $\lambda$, $\sigma_l$, and $\rho$ used in the data generating process.

\subsubsection{Evaluation}
We estimate the FWER for $p\leq 0.05$, which has a nominal rate of 5\%, and also estimate coverage of 95\% confidence sets. We also estimate the mean 95\% confidence interval width for each parameter $\delta$ to compare the efficiency of the procedures. 

\subsection{Results}

\begin{figure}
    \centering
    \includegraphics[width=\textwidth]{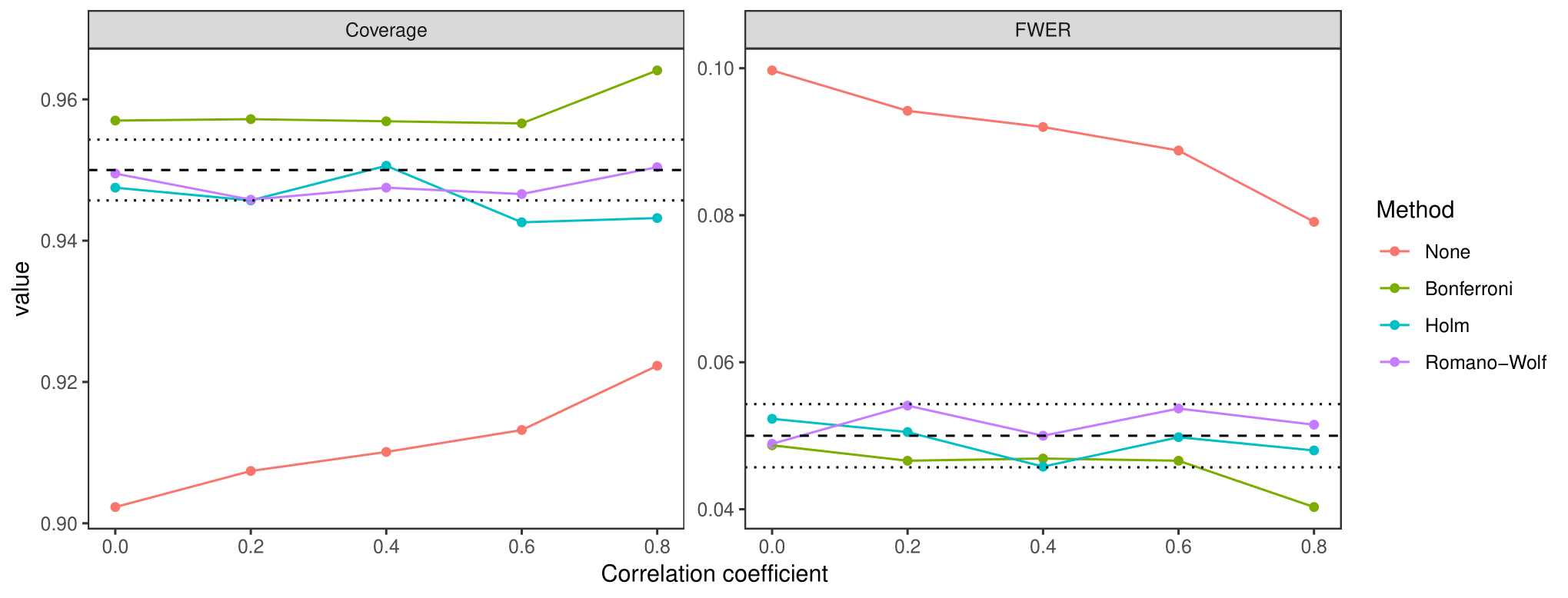}
    \caption{Family wise error rate and coverage under model (1) for four methods with different levels of the correlation coefficient $\rho$. The dashed line shows the nominal rates and the dotted lines approximate Monte Carlo confidence intervals. `None' refers to no correction.}
    \label{fig:rho}
\end{figure}

\begin{figure}
    \centering
    \includegraphics[width=0.7\textwidth]{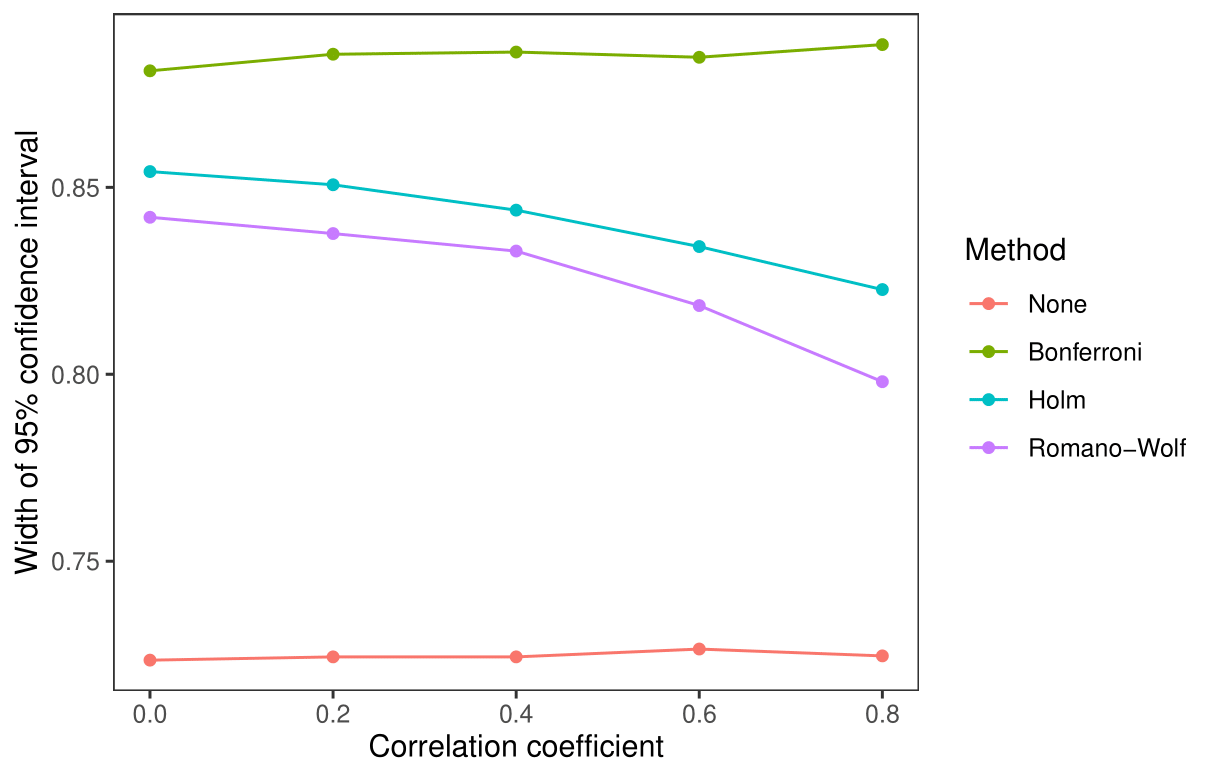}
    \caption{95\% Confidence interval width model (1) for four methods with different levels of the correlation coefficient $\rho$. `None' refers to no correction.}
    \label{fig:rho2}
\end{figure}

Figure \ref{fig:rho} shows the family wise error rates and coverage from model (1) with the permutation-based methods for different levels of the correlation coefficient. We exclude the `naive' approach from these plots as it has non-nominal marginal type I error and coverage without correction (see below). All three corrections ensured nominal error rates at lower levels of correlation ($\rho \leq 0.6$), however at higher levels of correlation Bonferroni was conservative. Without correction, the FWER declined as the correlation increased but was still approximately 0.08 at $\rho = 0.8$. Only the Romano-Wolf and Holm methods ensured nominal family wise coverage at any level of correlation. Figure \ref{fig:rho2} shows the 95\% confidence interval width for the four methods for the same model. For the two methods with nominal or near nominal error rates (Romano-Wolf and Holm), Romano-Wolf was moderately more efficient with narrower confidence intervals. The other methods displayed approximately constant confidence interval widths, with their respective widths reflecting the coverage results. 

Table \ref{tab:sims} reports the results from model (2). Under all tested conditions the FWER was approximately nominal in all scenarios for all multiple testing corrections when both parameters were zero. However, when only one parameter was zero, Bonferroni was conservative as expected with a FWER $\approx 0.025$ at $\alpha = 0.05$ which was also reflected in coverage being greater than the nominal rate. Romano-Wolf and Holm had nominal rates in all scenarios. Confidence interval width followed the same pattern as model (1) with Romano-Wolf generally being more efficient. Use of the weighted test statistic did not make much difference qualitatively with some confidence intervals larger and some smaller. Without correction, using a permutation test approach resulted in a FWER of $\approx ~0.10$ when there were two true null hypotheses, as expected. Using the naive output of \texttt{lme4} resulted in even worse performance due to the small sample bias in the test statistics, also as expected \citep{Leyrat2018, Watson2021}, with FWER around 30-50\% higher. Table \ref{tab:sims3} reports the results from the two outcome trial simulations with a larger 20 clusters per arm. The same pattern is observed as the smaller two-arm experiments, but the small sample bias using the naive method is reduced. To illustate the computational efficiency of the procedure, a single run of the function to derive p-values and confidence sets took between 1 and 10 seconds depending on the number of outcomes and size and number of the clusters.

\begin{table}
   \centering
    \begin{tabular}{lcc|cccc}
    \toprule
        \textbf{Method} & \textbf{Test statistic}  & $\delta$ & \textbf{FWER} & \textbf{Coverage} & \multicolumn{2}{c}{\textbf{CI width}}  \\
        &&&&& $\delta_1$ & $\delta_2$ \\
        \midrule
        None (naive) & -  & \multirow{9}{*}{(0,0)} & 0.158 & 0.844 & 0.653 & 0.497 \\
        \multirow{2}{*}{None (permutation)} & Unweighted &    & 0.099 & 0.902 & 0.725 & 0.605 \\
         & Weighted &  & 0.102 & 0.903 & 0.726 & 0.599 \\
        \multirow{2}{*}{Bonferroni} & Unweighted &    & \underline{0.048} & 0.958 & 0.881 & 0.746 \\
         & Weighted &  & \underline{0.051} & \underline{0.954} & 0.900 & 0.761 \\
        \multirow{2}{*}{Holm} & Unweighted &   & \underline{0.051} & \underline{0.947} & 0.855 & 0.716 \\
         &Weighted &     & 0.056 & 0.942 & 0.869 & 0.710 \\
        \multirow{2}{*}{Romano-Wolf} & Unweighted &    & \underline{0.053} & \underline{0.948} & 0.841 & 0.708 \\
         & Weighted &    & \underline{0.049} & \underline{0.947} & 0.824 & 0.720 \\
        \midrule
        None (naive) & -  &  \multirow{9}{*}{(0,0.5)} & 0.067 & 0.845 & 0.654 & 0.484 \\
        \multirow{2}{*}{None (permutation)} & Unweighted &   & \underline{0.048} & 0.914 & 0.722 & 0.637 \\
         & Weighted &   & \underline{0.049} & 0.916 & 0.727 & 0.650 \\
        \multirow{2}{*}{Bonferroni} & Unweighted & & 0.026 & 0.960 & 0.885 & 0.789 \\
         & Weighted & & 0.022 & 0.962 & 0.902 & 0.831 \\
        \multirow{2}{*}{Holm} & Unweighted &  & \underline{0.049} & \underline{0.954} & 0.851 & 0.754 \\
         &Weighted &  & 0.045 & \underline{0.953} & 0.870 & 0.773\\
        \multirow{2}{*}{Romano-Wolf} & Unweighted &  & \underline{0.048} & 0.957 & 0.839 & 0.740 \\
         & Weighted &  & \underline{0.051} & \underline{0.947} & 0.819 & 0.796 \\
         \bottomrule
    \end{tabular}
    \caption{Results of simulation experiments with two outcomes, seven clusters per arm, and with 10,000 iterations each. Each iteration used 1,000 permutations for the permutation test and 2,000 iterations in each of the lower and upper confidence interval search processes. \underline{underlined} results for FWER and coverage show those within approximated 95\% Monte Carlo confidence interval of the nominal value.}
    \label{tab:sims}
\end{table}

\begin{table}
   \centering
    \begin{tabular}{lcc|cccc}
    \toprule
        \textbf{Method} & \textbf{Test statistic}  & $\delta$ & \textbf{FWER} & \textbf{Coverage} & \multicolumn{2}{c}{\textbf{CI width}}  \\
        &&&&& $\delta_1$ & $\delta_2$ \\
        \midrule
        None (naive) & - &  \multirow{9}{*}{(0,0)} & 0.125 & 0.878 & 0.573 & 0.416 \\
        \multirow{2}{*}{None (permutation)} & Unweighted  &  & 0.095 & 0.908 & 0.597 & 0.457 \\
         & Weighted  &  & 0.099 & 0.901 & 0.600 & 0.466 \\
        \multirow{2}{*}{Bonferroni} & Unweighted  &  & \underline{0.053} & \underline{0.953} & 0.706 & 0.543 \\
         & Weighted  &  & 0.046 & 0.959 & 0.714 & 0.566 \\
        \multirow{2}{*}{Holm} & Unweighted   &  & \underline{0.053} & \underline{0.948} & 0.693 & 0.529 \\
         &Weighted  &  & \underline{0.048} & \underline{0.950} & 0.700 & 0.544 \\
        \multirow{2}{*}{Romano-Wolf} & Unweighted  & & \underline{0.052} & \underline{0.951} & 0.686 & 0.527 \\
         & Weighted  & & \underline{0.047} & \underline{0.952} & 0.687 & 0.539 \\
        \midrule
        None (naive) & -  & \multirow{9}{*}{(0,0.5)} & 0.057 & 0.878 & 0.871 & 0.399 \\
        \multirow{2}{*}{None (permutation)} & Unweighted  &  & \underline{0.049} & 0.915 & 0.597 & 0.469 \\
         & Weighted  &  & \underline{0.052} & 0.927 & 0.600 & 0.561 \\
        \multirow{2}{*}{Bonferroni} & Unweighted  &  & 0.023 & 0.961 & 0.707 & 0.558 \\
         & Weighted  & & 0.026 & 0.969 & 0.715 & 0.690 \\
        \multirow{2}{*}{Holm } & Unweighted  &  & \underline{0.053} & 0.935 & 0.693 & 0.546 \\
        &Weighted  & & \underline{0.049} & 0.963 & 0.698 & 0.660 \\
        \multirow{2}{*}{Romano-Wolf} & Unweighted  & & \underline{0.049} & 0.956 & 0.685 & 0.542 \\
         & Weighted  & & \underline{0.048} & \underline{0.960} & 0.679 & 0.655 \\
         \bottomrule
    \end{tabular}
    \caption{Results of simulation experiments with two outcomes, 14 clusters per arm, 10 individuals per cluster, and with 10,000 iterations each. Each iteration used 1,000 permutations for the permutation test and 2,000 iterations in each of the lower and upper confidence interval search processes. \underline{underlined} results for FWER and coverage show those within approximated 95\% Monte Carlo confidence interval of the nominal value.}
    \label{tab:sims3}
\end{table}

Table \ref{tab:sims2} shows the results from the three outcome simulations with baseline measures. Despite the more complex covariance structure and imbalance in the number of observations between control and treatment conditions, Holm and Romano-Wolf maintained nominal FWER and coverage. Again, Romano-Wolf was the most efficient correction. Its confidence intervals were between 10 and 50\% larger than the uncorrected results. We also note that the uncorrected approach maintain marginally nominal rates for each univariate outcome in all scenarios.

\begin{table}
\centering
    \begin{tabular}{lcc|ccccc}
    \toprule
        \textbf{Method} & \textbf{Test statistic}  & $\delta$ & \textbf{FWER} & \textbf{Coverage} & \multicolumn{3}{c}{\textbf{CI width}}  \\
        &&&&&$\delta_1$ & $\delta_2$ & $\delta_3$ \\
        \midrule
        \multirow{2}{*}{None (permutation)} & Unweighted  & \multirow{8}{*}{(0,0,0)} & 0.129 & 0.876 & 0.872 & 1.199 & 0.700 \\
         & Weighted &    & 0.148 & 0.848 & 1.033 & 1.201 & 0.767 \\
        \multirow{2}{*}{Bonferroni} & Unweighted & & \underline{0.046} & \underline{0.954} & 1.332 & 1.714 & 1.050 \\
         & Weighted &  & \underline{0.047} & 0.962 & 1.834 & 1.709 & 1.141 \\
        \multirow{2}{*}{Holm} & Unweighted &  & \underline{0.048} & \underline{0.946} & 1.202 & 1.652 & 1.092 \\
         & Weighted &  & \underline{0.048} & 0.942 & 1.554 & 1.612 & 1.101 \\
        \multirow{2}{*}{Romano-Wolf} & Unweighted & & \underline{0.052} & 0.956 & 1.014 & 1.633 & 0.986 \\
         & Weighted & & \underline{0.049} & \underline{0.954} & 1.611 & 1.498 & 0.997 \\
        \midrule
        \multirow{2}{*}{None (permutation)} & Unweighted & \multirow{8}{*}{(0,0.5,0)} & 0.093 & 0.869 & 0.783 & 1.271 & 0.705  \\
         & Weighted &  & 0.093 & 0.852 & 0.996 & 1.282 & 0.724 \\
        \multirow{2}{*}{Bonferroni} & Unweighted &  & 0.038 & 0.940 & 1.303 & 1.895 & 1.043 \\
         & Weighted &  & 0.032 & 0.942 & 1.824 & 1.834 & 1.134  \\
        \multirow{2}{*}{Holm}& Unweighted &  & 0.055 & 0.939 & 1.234 & 1.730 & 1.090 \\
         & Weighted & & 0.045 & 0.949 & 1.593 & 1.646 & 1.105 \\
        \multirow{2}{*}{Romano-Wolf } & Unweighted &  & \underline{0.048} & \underline{0.953} & 1.000 & 1.820 & 1.008 \\
        & Weighted &  & \underline{0.049} & \underline{0.954} & 1.678 & 1.548 & 1.031 \\
         \bottomrule
    \end{tabular}
    \caption{Results of simulation experiments for the parallel cluster trial with baseline measures (3). Each iteration used 1,000 permutations for the permutation test and 2,000 iterations in each of the lower and upper confidence interval search processes.}
    \label{tab:sims2}
\end{table}


\section{Applied example}
To provide a real-world example of the the use of the methods proposed in this article, we re-analyse a cluster randomised trial of a financial incentive to improve workplace health and wellbeing in small and medium sized enterprises (SME) in the United Kingdom. The original trial was relatively complex and included four trial arms with pre- and post-intervention observations comprising a standard control condition (no incentive), two treatment conditions (high and low incentive), and a second control arm with no baseline measures also with no incentive. The trial enrolled 152 clusters (SMEs), which were randomly allocated in an equal ratio to each of the trial arms; 100 SMEs completed the trial. Up to 15 employees were sampled and interviewed from each cluster. The full protocol is published elsewhere\citep{ThriveatWorkWellbeingProgrammeCollaboration2019a} (at the time of writing the results from the trial are under review).

\subsection{Outcomes}
A single primary outcome was specified in the protocol, which was the question ``Does your employer take positive action on health and wellbeing?''. However, given the potential lack of insight it might provide into the functioning of the intervention, several secondary outcomes were specified to capture the ``causal chain'' between intervention and employee health and wellbeing. For each of three separate health categories (mental, musculoskeletal, and lifestyle health) employees were asked:
\begin{enumerate}
    \item whether the employer provided information in this area;
    \item whether the employer had provided activities and services in this area;
    \item whether the employee had made a conscious effort to improve in this area;
    \item whether the employee had attended any groups or activities in this area at work;
    \item whether the employee had attended any groups or activities in this area outside of work.
\end{enumerate}
for a total of 15 outcomes.

\subsection{Re-analysis}
The original analysis of the trial took a Bayesian approach. The Frequentist 
re-analysis we conduct here is principally for illustrative purposes, and so we only take a subset of the data and simplify some of the outcomes. In particular, we take only the main control arm and the high incentive intervention arm to estimate the effect of the high incentive. We focus on the set of secondary outcomes listed above, which we collapse into five separate outcomes; whether the employer provided information across \textit{all} three health areas, and then whether there was a positive response for \textit{any} of the health areas for the remaining outcomes, for a total of five outcomes. All outcomes are modelled using a Bernoulli-logistic regression model, following the notation above, with $t=0$ for baseline and $t=1$ for post-intervention:
\begin{equation}
    Y_{k,ict} \sim \text{Bernoulli}(\text{logit}(\mu_{0,k} + \delta_k D_{ct} +  \theta_{k,c}  +  \theta_{k,ct}))
\end{equation}
We used 4,000 permutation test iterations and 10,000 steps in the confidence interval search procedure. For illustration, this re-analysis took eight minutes on a desktop PC with Intel Core i7-9700K with 16GB RAM and Windows 10.

\subsection{Results}
\begin{table*}
    \centering
    \begin{tabular}{lccccc}
    \toprule
    \textbf{Outcome} &\textbf{Statistic} & \textbf{None (Naive)} & \textbf{Bonferroni} & \textbf{Holm} &  \textbf{R-W} \\
    \midrule
    \multirow{5}{*}{\parbox{4cm}{Employer provided information}}
        & Estimate & \multicolumn{4}{c}{2.91}  \\
        & 95\% CI (Unweighted) & [1.98, 3.97] & [0.14, 2.96] & [-3.33, 3.56] & [0.34, 2.96]\\
        & \textit{p}-value (Unweighted)& 0.03 & 0.03 & 0.04 & 0.01 \\
        & 95\% CI (Weighted) &  & [-3.82,3.56] & [-1.97, 3.56] & [0.28, 2.96]\\
        & \textit{p}-value (Weighted) & & 0.02 & 0.02 & <0.01 \\
        \midrule
       \multirow{5}{*}{\parbox{4cm}{Employer provided activities}}  & Estimate & \multicolumn{4}{c}{2.11} \\
        & 95\% CI (Unweighted)& [1.31, 2.99]& [-0.11, 3.18] & [-0.19, 2.90] &[-0.11, 3.22]\\
        & \textit{p}-value (Unweighted) & $<$0.01 & 0.21 & 0.15 &  0.05 \\
        & 95\% CI (Weighted)& & [-3.53, 3.02] & [-3.46, 2.71] &[-0.16, 2.95]\\
        & \textit{p}-value (Weighted)&  & 0.21 & 0.13 &  0.04 \\
        \midrule
        \multirow{5}{*}{\parbox{4cm}{Employee made a conscious effort}} & Estimate & \multicolumn{4}{c}{0.22} \\
        & 95\% CI (Unweighted)& [-0.33, 0.77] & [-0.89, 0.98] &  [-0.65, 0.95] & [-0.77, 1.45] \\
        & \textit{p}-value (Unweighted)&  0.44  & 1.00 & 0.38 & 0.37\\
        & 95\% CI (Weighted)&  & [-0.68, 1.11] &  [-0.59, 1.39] & [-0.84, 1.45] \\
         & \textit{p}-value (Weighted)&   & 1.00 & 0.38 & 0.36\\
         \midrule
         \multirow{5}{*}{\parbox{4cm}{Employee took part at work}} & Estimate & \multicolumn{4}{c}{1.13}\\
         & 95\% CI (Unweighted)& [0.50, 1.75] & [-0.37, 1.72]& [-0.45, 1.79] & [-0.39, 1.85] \\
         & \textit{p}-value (Unweighted)& $<$0.01 & 1.00 & 0.88 & 0.27\\
        & 95\% CI (Weighted)&  & [-3.15, 1.75] & [-0.47, 1.73] & [-0.43, 1.90] \\
         & \textit{p}-value (Weighted)& & 1.00 & 0.84 & 0.29\\
         \midrule
        \multirow{5}{*}{\parbox{4cm}{Employee took part outside work}}&Estimate & \multicolumn{4}{c}{0.27}\\
         & 95\% CI (Unweighted)& [-0.06, 0.61] & [-0.09, 0.95] & [-0.08, 0.97] & [-0.69, 0.83]\\
          & \textit{p}-value (Unweighted)& 0.11 & 0.34 & 0.17 & 0.18\\
          & 95\% CI (Weighted)& & [-0.13, 0.99] & [-0.17, 0.96] & [-0.70, 0.83]\\
          & \textit{p}-value (Weighted)& & 0.34 & 0.16 & 0.17\\
         \bottomrule
    \end{tabular}
    \caption{Results from re-analysis of the workplace wellbeing trial. Results are log odds-ratios, 95\% confidence intervals, and p-values. Permutation test p-values used 4,000 iterations, and the confidence interval search procedure used 10,000 steps for Bonferroni, Holm, and Romano-Wolf (RW) methods. The `None (Naive)' method refers to a model-based analysis using \texttt{lme4} with no multiple testing correction.}
    \label{tab:reanal}
\end{table*}

\begin{figure}
    \centering
    \includegraphics[width=\textwidth]{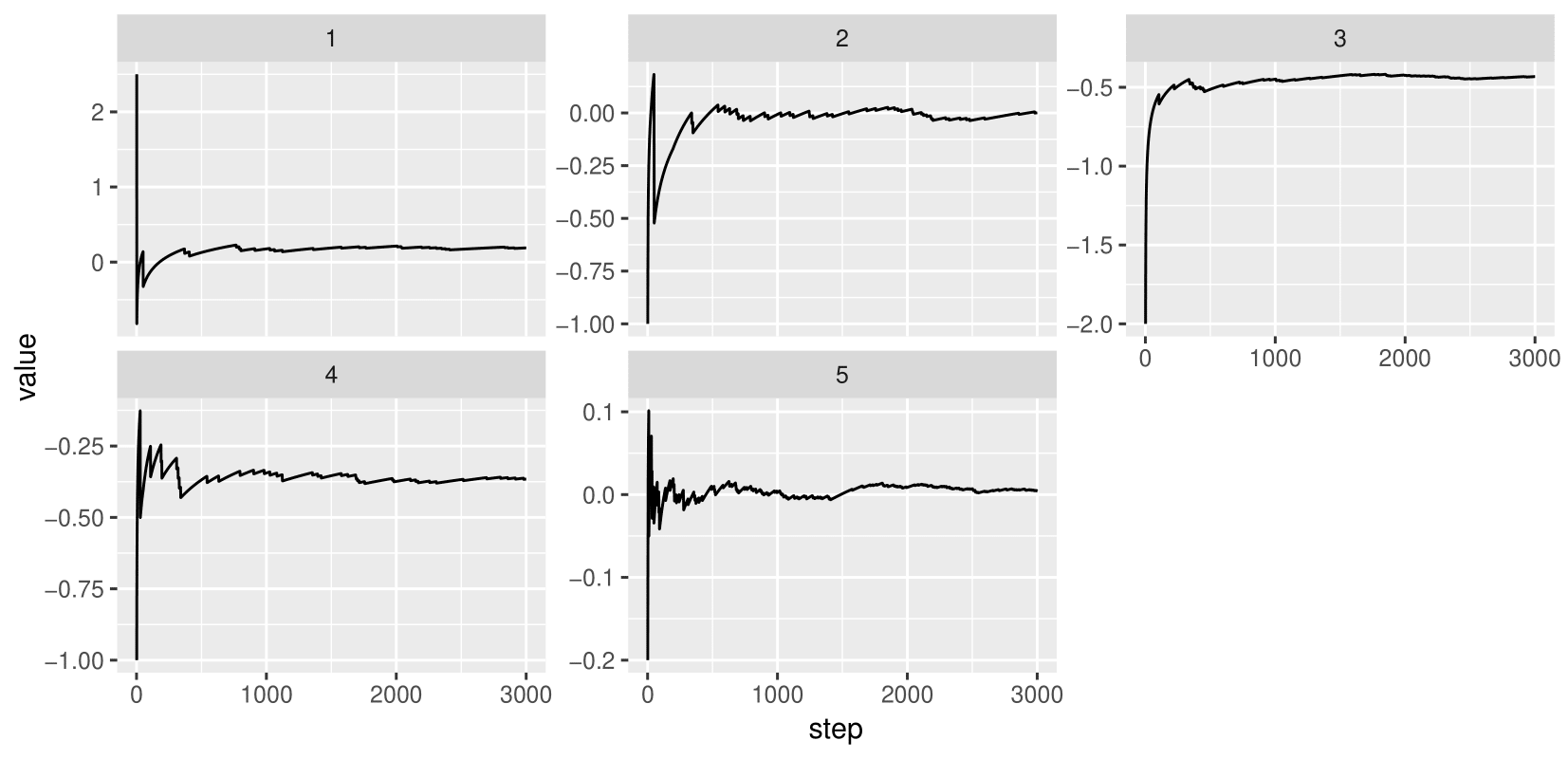}
    \caption{Example of the confidence interval search for the lower confidence interval limit using the Holm correction for the cluster trial example.}
    \label{fig:confint}
\end{figure}

Table \ref{tab:reanal} shows the results of an analysis using the naive method (a model-based analysis using \texttt{lme4} with no multiple testing correction), alongside `corrected' results using the Bonferroni, Holm, and Romano-Wolf methods. We first note that the convergence of the confidence interval search procedure was highly sensitive to the starting values. The algorithm could take a long time to find the right part of the parameter space, particularly since the search distance decays with the number of iterations. Convergence can be assessed graphically; the chain `osciallates' around a value at convergence compared to continuous gradual climbing or descending, Figure \ref{fig:confint} shows an example. 

We make several observations about the results. The uncorrected analysis would suggest there is likely good evidence that the intervention improved employer provision of information and activities and services, and increased employee taking part at work. However, this conclusion might contradict our understanding of the causal processes since it would seem contradictory for employees to make more effort but not report making more effort. The results corrected for multiple testing using Romano-Wolf appear to be more consistent in that employers appeared to make more effort but the employees did not take up the new services with small and negative effects now shown to be compatible with the data for the latter three outcomes. The effect of the intervention is also more uncertain than suggested by the uncorrected confidence intervals. In particular, the confidence intervals under the corrected methods, which are based on exact permutation tests, are not symmetric for several outcomes, unlike under the uncorrected approach. So, smaller effect sizes, particularly for the first two outcomes, are more plausible than the uncorrected method would suggest. 

\section{Discussion}
We have proposed how one can estimate Frequentist statistics for cluster randomised trials with multiple outcomes that control for the FWER and coverage of simultaneous confidence intervals. These methods also apply generally in any scenario where multiple tests from GLMMs are used. Where a correction for multiple testing is desired in a cluster trial setting, the Romano-Wolf approach would be recommended as it maintains nominal rates in a variety of scenarios including with differing levels of between-outcome correlation, cluster and individual sample sizes, and covariance structures, it is also more efficient than the alternatives. Where a multiple testing correction is not desired, permutation-based methods are likely to provide marginally nominal error rates and so are also recommended when other methods may exhibit biases. We also compared a weighted test statistic based on the score statistic proposed by Romano and Wolf\citep{Romano2005}, but did not find this provided any obvious benefit over an unweighted sum of generalised residuals. We do note, however, that while these methods do provide the desired properties, many regulatory agencies, including the FDA, do not (yet) accept statistics derived from re-sampling based methods, which may limit their application. Researchers may also consider other methods if multiple testing corrections are required such as `intersection-union' testing.\citep{Yang2022}

There have been no previous comparisons of multiple testing corrections in the context of cluster randomised trials as far as we are aware, but our results generally reflect those from other settings. For example, Ozenne et al\citep{Ozenne2022} compared several multiple testing corrections for linear latent variable models, including a resampling-based procedure, although not Romano-Wolf. They showed this method maintained strong control of the FWER and was more efficient than Bonferroni. Vickerstaff et al\citep{Vickerstaff2019} considered the question for individual level randomised trials with a linear model, and suggested that Hommel's \citep{HOMMEL1988} and Hochberg's \citep{HOCHBERG1988} methods were marginally more efficient than Bonferroni or Holm, but they did not include a permutation-based procedure, not non-linear models. Alberton et al\citep{Alberton2020} also shows permuation-based methods to outperform other corrections in the context of analysing brain imaging data.

We have examined methods from a range of previous work including: permutation tests for cluster trials,\citep{Gail1996a,Gallis2018} univariate methods for corrections for multiple testing that use permutation tests,\citep{Romano2005,Romano2005a,Romano2016} and procedures for estimating confidence interval limits based on permutation tests.\citep{Garthwaite1992,Garthwaite1996,Rabideau2021,Rabideau2021b} Altogether the proposed methods can deal with several issues that are common to cluster randomised trials as they allow for multiple outcomes, they can incorporate other features such as restricted randomisation methods, which are often used in trials with a small number of clusters. Watson et al,\citep{Watson2021} Li et al,\citep{Li2016,Li2017} and Zhou et al\citep{Zhou2022} discuss permutation tests with restricted randomisation methods. Permutation-based methods provide exact inference when there are a small number of clusters, which can lead to non-nominal error rates of standard test procedures and hence confidence intervals with non-nominal coverage. Several small-sample corrections exist that can provide nominal error rates with a small number of clusters,\citep{Watson2021,Leyrat2018} however there is no obvious way these would be incorporated efficiently into a multiple testing procedure. After conducting the analyses presented in this article, an updated and more efficient version of the confidence interval search procedure was brought to our attention \citep{Garthwaite2009}. This method improves the efficiency of the search procedure, and requires fewer steps by making larger steps on average, although would not affect the results presented here. We aim to incorporate the algorithm in our R package implementing these methods (\texttt{crctStepdown}).

The tools developed for this article can be incorporated at the design stage of a cluster trial to determine power using simulation-based approaches. These methods are useful for the analysis of cluster trials with multiple outcomes and the treatment effect parameters from the linear predictors of multiple univariate models, however, it is not clear how or if they could be applied to cluster trials with multiple \textit{arms}. In multi-arm trials there may be one or more outcomes, but clusters may receive different `doses' or variants of the treatment. There are a variety of treatment effects and null hypotheses of interest including pairwise comparisons between arms and a global joint null, which can be estimated from a single univariate model with indicators for each arm.\citep{Watson2021,Zhou2022} Pairwise null hypotheses in these models do not make statements about the value of the treatment effects in arms outside the pair under comparison as it is left unspecified, so it is not obvious then how a permutation test could be conducted for the pairwise comparison that is invariant to randomised allocation. The multiple treatment effects of interest in a multi-arm study clearly fall in the realm of multiple testing. Nevertheless, we believe the methods proposed in this article will be a useful tool for the analysis of cluster randomised trials in many cases.




\bibliography{multiple}





\end{document}